\documentclass[aps,twocolumn,floatfix]{revtex4-2}
\usepackage{amssymb}
\usepackage{natbib}
\usepackage{graphicx}
\usepackage{amsmath}
\usepackage[bookmarks = false,colorlinks = true,allcolors = blue]{hyperref}
\usepackage{bm}
\usepackage[all]{hypcap}
\usepackage{graphicx}
\usepackage{colortbl}
\usepackage{booktabs}
\usepackage{enumerate}
\usepackage{enumitem}
\usepackage{braket}
\usepackage{mathrsfs}
\usepackage{multirow}
\usepackage{upgreek}
\usepackage{textcomp}
\usepackage{MnSymbol}

\usepackage{lineno}
\usepackage{threeparttable}
\usepackage{romannum}
\usepackage{array}
\newcommand{\PreserveBackslash}[1]{\let\temp=\\#1\let\\=\temp}
\newcolumntype{C}[1]{>{\PreserveBackslash\centering}p{#1}}
\newcolumntype{R}[1]{>{\PreserveBackslash\raggedleft}p{#1}}
\newcolumntype{L}[1]{>{\PreserveBackslash\raggedright}p{#1}}

\begin{document}
\pagenumbering{arabic}
\title{Entangling metropolitan-distance separated quantum memories}

\author{Xi-Yu Luo$^{1,\,2,\,*}$}
\author{Yong Yu$^{1,\,2,\,*}$}
\author{Jian-Long Liu$^{1,\,2}$}
\author{Ming-Yang Zheng$^{3}$}
\author{Chao-Yang Wang$^{1,\,2}$}
\author{Bin Wang$^{1,\,2}$}
\author{Jun Li$^{1,\,2}$}
\author{Xiao Jiang$^{1,\,2}$}
\author{Xiu-Ping Xie$^{3}$}
\author{Qiang Zhang$^{1,\,2,\,3}$}
\author{Xiao-Hui Bao$^{1,\,2}$}
\author{Jian-Wei Pan$^{1,\,2}$}

\affiliation{$^1$Hefei National Laboratory for Physical Sciences at Microscale and Department
of Modern Physics, University of Science and Technology of China, Hefei,
Anhui 230026, China}
\affiliation{$^2$CAS Center for Excellence in Quantum Information and Quantum Physics, University of Science and Technology of China, Hefei, Anhui 230026, China}
\affiliation{$^3$Jinan Institute of Quantum Technology, Jinan, China.}
\affiliation{$^*$These two authors contributed equally to this work.}

\begin{abstract}
Quantum internet gives the promise of getting all quantum resources connected, and it will enable applications far beyond a localized scenario. A prototype is a network of quantum memories that are entangled and well separated. Previous realizations are limited in the distance. In this paper, we report the establishment of remote entanglement between two atomic quantum memories physically separated by 12.5 km directly in a metropolitan area. We create atom-photon entanglement in one node and send the photon to a second node for storage. We harness low-loss transmission through a field-deployed fiber of 20.5 km by making use of frequency down-conversion and up-conversion. The final memory-memory entanglement is verified to have a fidelity of 90\% via retrieving to photons. Our experiment paves the way to study quantum network applications in a practical scenario. 
\end{abstract}

\maketitle

Quantum memory is an essential element in a quantum network~\cite{kimble2008,wehner2018}, since it mediates the photonic qubit transmissions and the matter qubit manipulations. A prototype of quantum networks is the entanglement of well-separated quantum memories. Experimentally, two-node entanglement has been realized through various approaches, such as solid-state impurities~\cite{hensen2015}, quantum dots~\cite{delteil2016}, trapped ions~\cite{hucul2015} and neutral atoms~\cite{hofmann2012,daiss2021}, cold atomic ensembles~\cite{yu2020}, rare-earth ion ensembles~\cite{lago-rivera2021,liu2021a}. Extension to three nodes was also reported recently~\cite{jing2019,pompili2021}. Moving from these 
proof-of-principle experiments to a genuine quantum network in the metropolitan regime is not only indispensable for the promising applications (such as device-independent quantum key distribution, deterministic quantum teleportation~\cite{pfaff2014,langenfeld2021}, quantum repeater~\cite{briegel1998}, distributed quantum computing, entanglement-based clock synchronization~\cite{komar2014}), but also significantly meaningful for the test of quantum foundations~\cite{hensen2015,rosenfeld2017}. The longest separation between two quantum nodes so far has a mere distance of 1.3 km~\cite{hensen2015}. Extension to longer distance is facing a number of challenges. One major limiting issue is that the photon wavelength of most memories is not suitable for low-loss transmission in optical fibers; thus, an efficient and low-noise quantum frequency converter (QFC)~\cite{radnaev2010,maring2017,degreve2012,bock2018,ikuta2018,vanleent2019,tchebotareva2019,krutyanskiy2019} is required. Another issue is that the two nodes need to be fully independent, which raises experimental complexities~\cite{yu2020} involving remote phase synchronization, etc. In addition, the memory needs to be long-lived with a lifetime significantly longer than the fiber transmission delay. 

Here we resolve these issues by reporting the establishment of entanglement between two quantum memories that are fully independent and long-distance separated. We make use of two quantum memory nodes separated by 12.5~km physically and connected with optical fibers of 20.5~km. The memories are based on laser-cooled atomic ensembles, for which key capabilities have been realized already, such as sub-second storage~\cite{yang2016,wang2021}, efficient retrieval~\cite{yang2016,cho2016,wang2019,jing2019,cao2020,wang2021}, spatial~\cite{pu2017,chrapkiewicz2017} and temporal~\cite{heller2020} multiplexing. In one node, we design a scheme to directly generate entanglement of the atomic ensemble with a single-photon in the time-bin degree. The atomic coherence is prolonged via zeroing the spin-wave wavevector. By transmitting the photon to the other node and storing it, we entangle the two remote quantum memories. The transmitted photon is shifted in frequency from the Rubidium D$_1$ line to the O-Band in fiber-optic communication to minimize the transmission losses. Working in the time-bin degree not only enhances the robustness of long-distance transmission, but also simplifies the design of the frequency conversion modules significantly. The final memory-memory entanglement is verified via retrieving to photons, giving a fidelity of 90\%. Our experiment provides a platform to study 
quantum network applications~\cite{wehner2018} in a practical scenario and test quantum foundations over metropolitan distances.

\begin{figure*}[htbp]
	\centering
	\includegraphics[width=\textwidth]{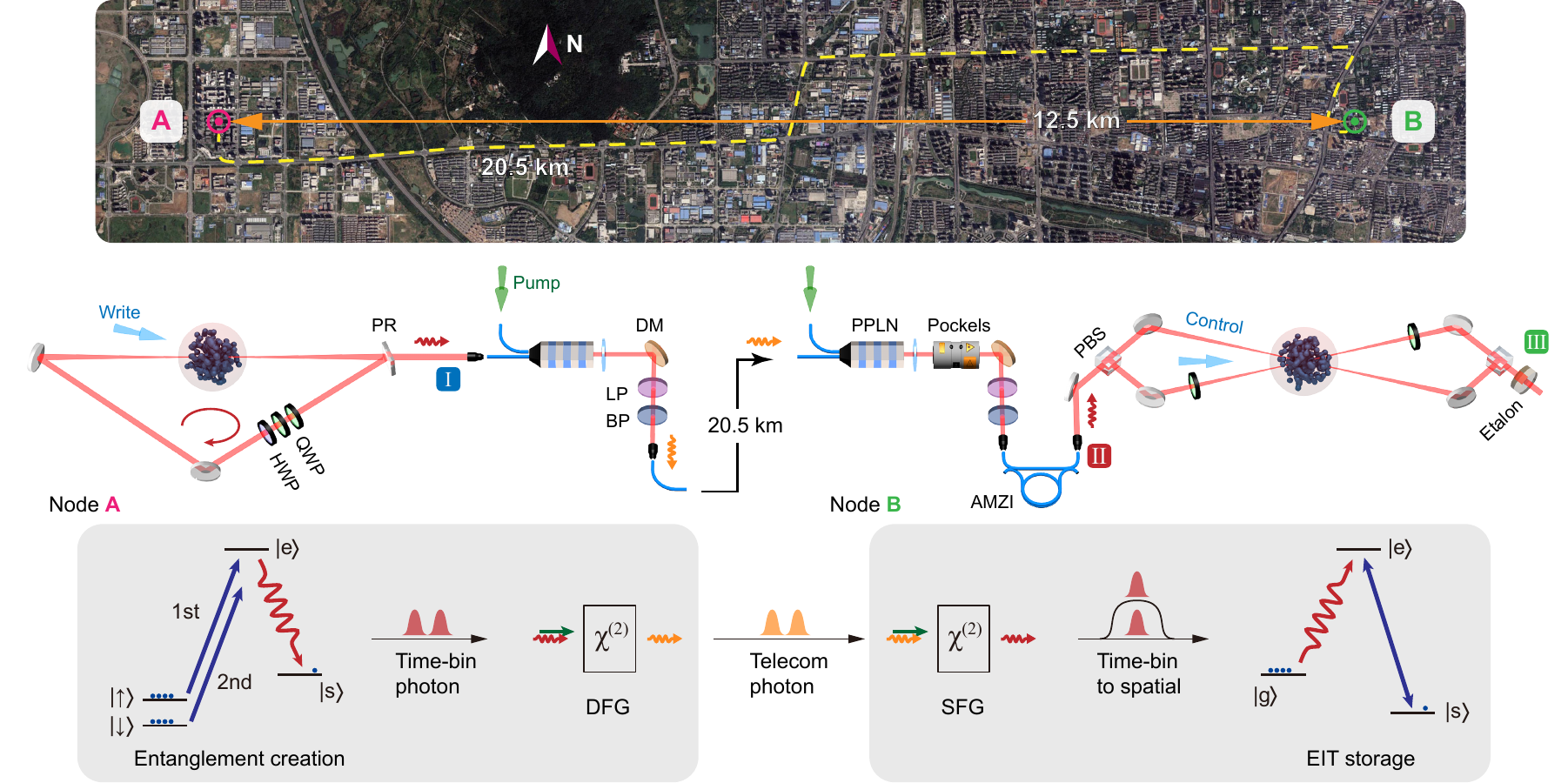}
	 \caption{\textbf{Experimental layout.} (\textbf{Top}) Bird's-eye view of the memory nodes (A and B) that are 12.5~km apart and connected with fibers of 20.5~km (17~km deployed together with a 3.5~km long fiber loop). Each node includes a setup with $^{87}$Rb atoms and a quantum frequency converter. Schematic (\textbf{Middle}) and corresponding level schemes (\textbf{Bottom}). In node A, a cavity-enhanced DLCZ-type setup is utilized to generate atom-photon entanglement. The write-out photon propagates along the clockwise mode of the cavity and emits from the partially reflective mirror (PR). The emitted photon is then converted to 1342 nm via DFG in a PPLN waveguide and transmitted to node B via the fiber channel. In node B, the photon is first converted back to 795~nm via SFG in another PPLN waveguide chip. A series of filters, including dichroic mirrors (DM), long-pass filters (LP), and band-pass filters (BP), are used to suppress the noise in each conversion module (The types and quantities shown in the figure are not in one-to-one correspondence with the actual situation). The photon's two time-bin modes are then converted to two spatial modes and stored in the atomic ensemble by applying a control pulse via the EIT mechanism. The time-bin to spatial conversion is achieved by a combination of a Pockels cell, an asymmetric Mach-Zehnder interferometer (AMZI), and a polarizing beamsplitter (PBS). An etalon in the readout path of the EIT quantum memory is for blocking the control pulse leakage. QWP and HWP are quarter- and half-wave plates, respectively. Marker \Romannum{1}, \Romannum{2} and \Romannum{3} indicate three checkpoints for the write-out photon during the characterization of the photon transfer and the EIT storage. Map data is from Google and Maxar Technologies.}
	\label{fig:setup}
\end{figure*}

The layout of our experiment is shown in Fig.~\ref{fig:setup}. It comprises two nodes distantly separated in Hefei city, labeled as A and B, and several field-deployed optical fibers linking them. We start by generating atom-photon time-bin entanglement in node A using an improved version of the Duan–Lukin–Cirac–Zoller (DLCZ) scheme~\cite{duan2001}. A laser-cooled $^{87}$Rb atomic ensemble are initially prepared as a mixture of $\ket{\downarrow}$ and $\ket{\uparrow}$, two Zeeman levels of the lowest atomic hyperfine ground state $\ket{g} \equiv \ket{5 S_{1/2}, F=1}$ with the magnetic quantum number $m_F=-1$ and $+1$, respectively (see Supplementary Material). Two write pulses with orthogonal polarizations drive spontaneous Raman scattering from $\ket{\downarrow}$ and $\ket{\uparrow}$ in sequence. In a small probability $\chi$, a write-out photon is scattered along the ring cavity mode through the early process along with a spin-wave $\ket{\Downarrow}=\sum_je^{i\Delta \vec{k}\cdot \vec{r}_j}\ket{\downarrow\dots s_j \dots\downarrow}$, or through the late process along with another spin-wave $\ket{\Uparrow}=\sum_je^{i\Delta \vec{k}\cdot \vec{r}_j}\ket{\uparrow\dots s_j \dots\uparrow}$, where $\Delta \vec{k}$ is the wavevector difference of write beam and write-out photon, also known as the wavevector of the spin-wave. These two spin-wave states form an atomic qubit. Ensuring coherence between two write pulses, we generate a maximally entangled atom-photon state
\begin{equation}\label{eq:apEnt}
	\ket{\Psi_{\mathrm{ap}}} =\frac{1}{\sqrt{2}}\left(\ket{\Downarrow}\ket{E}+e^{-i\varphi(t)}\ket{\Uparrow}\ket{L}\right), \nonumber
\end{equation}
where $E$ and $L$ denote the write-out photon's early and late time-bin mode, respectively. A time-dependent phase $\varphi(t)=\mu_B Bt/\hbar+\varphi_0$ is involved due to the Zeeman splitting induced by a bias magnetic field $B$, where $\mu_B$ is the Bohr magneton, $\hbar$ is the reduced Planck constant, and $t$ is the evolution time. Besides being intrinsically appropriate for long-distance transmission, the time-bin encoded photon is favorable for the following QFC process for avoiding polarization selection of nonlinear process. Two spin-wave modes can be efficiently retrieved on-demand as two orthogonal polarization photon modes for measurement. Write-out photons are next down-converted from Rubidium resonance to telecom O-Band to minimize the transmission attenuation. This is achieved by the difference frequency generation (DFG) process in a periodically poled lithium niobate (PPLN) waveguide chip with the help of a strong 1950~nm pump laser. 

\begin{figure*}[htbp]
	\centering
	\includegraphics[width=0.8\textwidth]{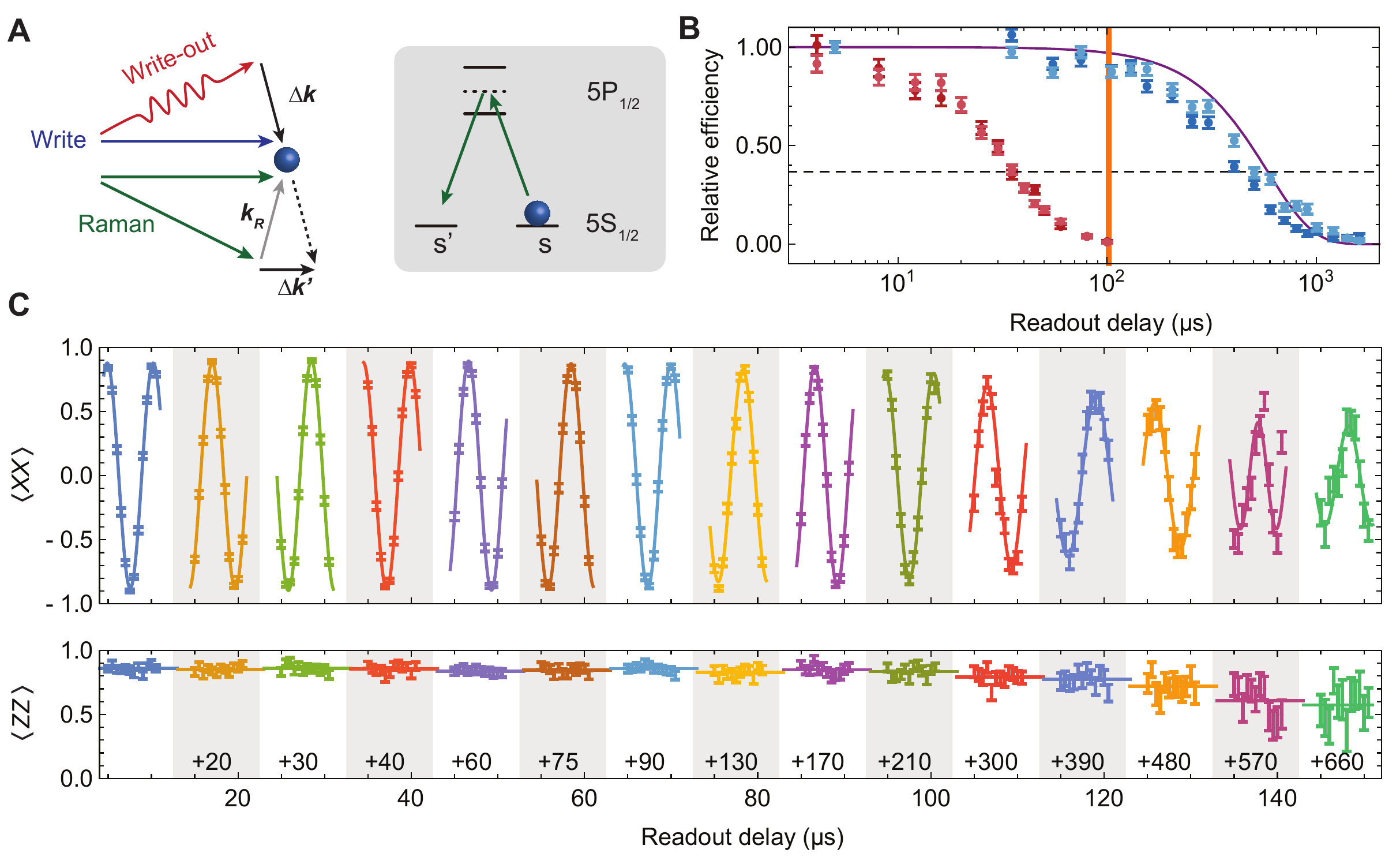}
	 \caption{\textbf{Benchmarking the DLCZ quantum memory.} (\textbf{A}) Scheme of the spin-wave freezing. After the spin-wave was created, a Raman $\pi$-pulse which couples the $\ket{s}$ to $\ket{s'}$ transition via a two-photon Raman process is applied. The Raman process introduces a new momentum $\hbar\vec{k}_R$ to the $\ket{s}/\ket{s'}$ state atom, resulting in an altered spin wave with the wavevector $\Delta \vec{k}'=\Delta\vec{k}+\vec{k}_R\approx 0$. The Raman $\pi$ pulse is applied again before the readout to recover the wavevector. (\textbf{B}) Relative retrieval efficiency as a function of the readout delay. The data for two spin-wave states are dark and light blue for the spin-wave freezing case and dark and light red for the spin-wave freezing free case. The purple curve is the theoretical expectation of the spin-wave freezing (see Supplementary Material). The orange bar indicates the entanglement distribution time in the experiment. (\textbf{C}) Measured atom-photon correlations at $\chi=5.4\%$ as a function of the readout delay. The oscillation of $\braket{XX}$ originates from the time evolution of the atomic phase induced by the magnetic field. Error bars indicate one standard deviation of the photon-counting statistics.
	 }
	\label{fig:local}
\end{figure*}

At node B, the telecom photon is up-converted back to 795~nm with the help of the sum-frequency generation (SFG) process in another PPLN waveguide chip to match the Rubidium resonance. Before the SFG process, we compensate the fiber-induced polarization drift by continously adjusting an electrical polarization controller~\cite{yu2020}. In this node, another laser-cooled $^{87}$Rb atomic ensemble initialized in $\ket{a}\equiv\ket{5S_{1/2},F=2,m_F=+2}$ serves as the quantum memory. By applying a strong control field coupling a stable state $\ket{b}\equiv\ket{5S_{1/2},F=1,m_F=0}$ with an excited state $\ket{c}\equiv\ket{5P_{1/2},F=1,m_F=+1}$, the input photon on resonance with $\ket{a}\leftrightarrow\ket{c}$ is mapped as a spin-wave $\sum_i\ket{a\dots c_i \dots a}$ through the electromagnetically induced transparent (EIT)~\cite{fleischhauer2005}. Two time-bin modes, $\ket{E}$ and $\ket{L}$, of the input photon, are transformed to two spatial modes up and down, respectively, and stored afterwards. In this way, we create a remote atom-atom entanglement
\begin{equation}\label{eq:aaEnt}
	\ket{\Psi_{\mathrm{aa}} }=\frac{1}{\sqrt{2}}\left(\ket{\Downarrow}\ket{U}+e^{-i\varphi(t)}\ket{\Uparrow}\ket{D}\right), \nonumber
\end{equation}
where $\ket{U}$ and $\ket{D}$ denote two spin-wave states of up and down spatial mode, respectively.

\begin{figure*}[htp]
	\centering
	\includegraphics[width=0.65\textwidth]{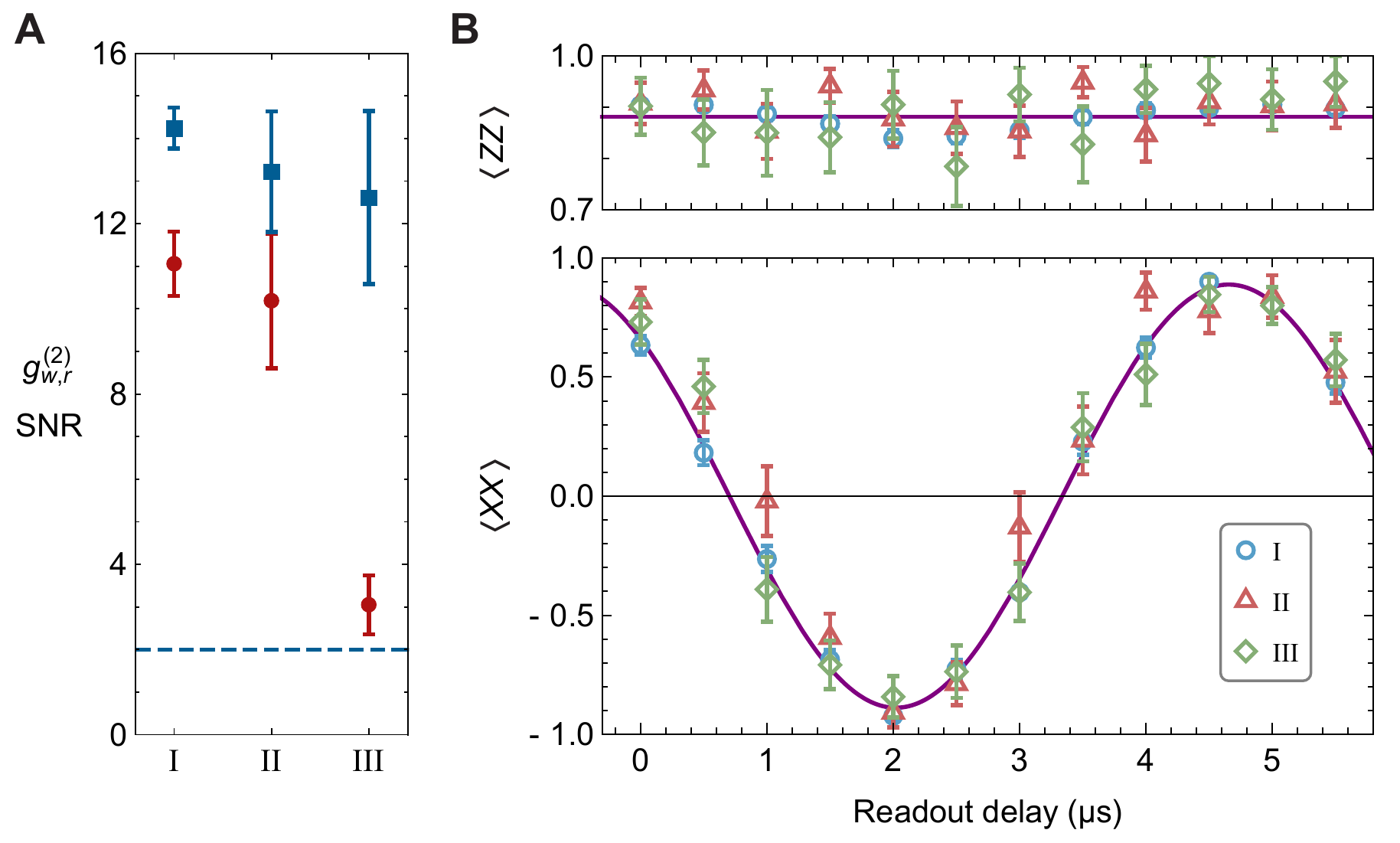}
	 \caption{\textbf{Characterization of the photon transfer and the EIT quantum memory.} (\textbf{A}) Red dots shows the measured SNR of the write-out photon before transfer (\Romannum{1}), after transfer (\Romannum{2}) and after EIT storage (\Romannum{3}). The normalized cross-correlation function $g^{(2)}_{w,r}$ between the write-out and the read-out photon at the corresponding three conditions are shown in blue squares. (\textbf{B}) Measured $\braket{ZZ}$ and $\braket{XX}$ correlations of $\ket{\Psi_{\text{ap}}}$ as a function of readout delay, i.e., the atomic phase evolution. Blue circles, red triangle, and green diamond refer to measuring the write-out photon at \Romannum{1}, \Romannum{2} and \Romannum{3}. The purple curve in each graph shows the fitting for data in case \Romannum{1}. Error bars indicate one standard deviation of the photon-counting statistics.}
	\label{fig:transfer}
\end{figure*}

Entanglement distribution between distant network nodes can be extremely time-consuming, which raises a critical demand for long-lived storage. In this experiment, it is essential that the quantum memory at node A can survive longer than the communication time of $103$~$\upmu$s between the two nodes. A dominant decoherence mechanism for spin-waves in the atomic ensemble is the thermal motion $\vec{r}'_j=\vec{v}_j t$ of atoms. It will introduce a random phase $\Delta\vec{k}\cdot\vec{r}_j'$ to each item and ruin the collective interference process during retrieval. Here, by making use of an auxiliary state $\ket{s^{\prime}} \equiv\ket{5^{2} S_{1 / 2}, F=2, m_{F}=-1}$, we coherently freeze the spin-wave~\cite{jiang2016} to minimize the thermal motion-induced decoherence as shown in Fig.~\ref{fig:local}A. Two Raman beams driving the atom from $\ket{s}$ to $\ket{s'}$ with a $\pi$ transition will kick the atom with a momentum of $\hbar\vec{k}_R$, leading to an engineered wavevector $\Delta\vec{k}'=\Delta \vec{k}+\vec{k}_{R}$, where $\vec{k}_R$ is the difference between the wavevector of two Raman beams. With an appropriate arrangement of Raman beams, $|\Delta\vec{k}'|$ is minimized to a near-zero value (see Supplementary Material). Note that the engineering process works for both spin-wave modes since their initial wavevectors are identical. We observed a $1/e$ lifetime of $416~\upmu$s and $517~\upmu$s for $\ket{\Downarrow}$ and $\ket{\Uparrow}$, respectively, as shown in Fig.~\ref{fig:local}B, both surpassing the entanglement distribution time in this experiment. A slightly longer lifetime of $\ket{\Uparrow}$ originates from its magnetic field insensitive clock-state energy level configuration during the spin-wave freezing.

As the atomic qubit in node A is phase-evolving under the bias magnetic field, its coherence relies on the stability of the magnetic field. Thus, we take passive and active measures to cancel magnetic noise (see Supplementary Material). To characterize the coherence of the atomic qubit, we prepare an entangled state as in Eq.~\ref{eq:apEnt} and measure the atom-photon correlation in $XX$ and $ZZ$ basis along with the storage time increase (Fig.~\ref{fig:local}C), where hereafter we use $X$, $Y$ and $Z$ as the shorthand for standard Pauli matrices $\sigma_x$, $\sigma_y$ and $\sigma_z$, respectively, for both atomic and photonic qubits. We can observe an oscillation under $XX$ basis caused by the time-evolving phase. By fitting the decrease of $\braket{ZZ}$ and the envelope of $\braket{XX}$, we deduce an average amplitude damping time $T_1 = 1.2$~ms and a phase damping time $T_2^{*}=856.7~\upmu$s for qubit storage. 

Next, we investigate the photon state transfer between two nodes and the EIT quantum memory in node B. The optimal end-to-end efficiency for DFG and SFG modules are 46\% and 45\%, respectively. Together with the 7.1~dB transmission losses of the fiber channel, we have a photon transfer efficiency from node A to B around $4\%$. As a comparison, the transmission efficiency of a 795~nm photon on the same fiber channel without frequency conversion will be on the order of $10^{-7}$. The EIT quantum memory storage efficiency, including mapping in and mapping out, is about $22\%$ and $25\%$ for two spatial modes, respectively. Noise photons introduced during the photon transfer and storage will lead to depolarization of the remote entanglement. To determine the noise strength, by setting $\chi=5.4\%$, we measure the signal-to-noise ratio (SNR) of the write-out photon at three different checkpoints, namely, before transfer at \Romannum{1}, after transfer at \Romannum{2}, and after storage and readout in the EIT quantum memory at \Romannum{3} (see labels in Fig.~\ref{fig:setup}) as shown in Fig.~\ref{fig:transfer}A. Thanks to the efficient noise filtering during QFC, the SNR hardly drops as the photon propagates from \Romannum{1} to \Romannum{2}. The obvious decrease of the SNR after EIT storage is due to a reduced signal strength that is approaching the dark counts level of the silicon-based single-photon detectors, which can be drastically mitigated by using superconducting nanowire single-photon detectors with ultra-low dark counts~\cite{shibata2015}. Nevertheless, the strong atom-photon correlation barely suffers from the noise. As also depicted in Fig.~\ref{fig:transfer}A, we observed a uniformed cross-correlation function between the write-out and the read-out photon $g^{(2)}_{w,r}$ of $13.2\pm 1.4$ and $12.6\pm 2.0$ when measuring the write-out photon at \Romannum{2} and \Romannum{3}, respectively, which are slightly lower than the initial value $14.2\pm 0.5$ when measuring the write-out photon at \Romannum{1}. We also measured the atom-photon correlation under different Pauli basis at these checkpoints (Fig.~\ref{fig:transfer}B). When sweeping the readout delay, i.e., the entanglement phase $\varphi(t)$, strong and near-identical correlation variation for $\braket{XX}$ and $\braket{ZZ}$ for the three checkpoints reveals a high-quality photon transfer and EIT storage. The atomic state for all cases is measured within 5~$\upmu$s ($\ll T_1,\, T_2^*$) after the entanglement was created to get rid of the influence from time-dependent dephasing.

\begin{figure}[htp]
	\centering
	\includegraphics[width=0.9\columnwidth]{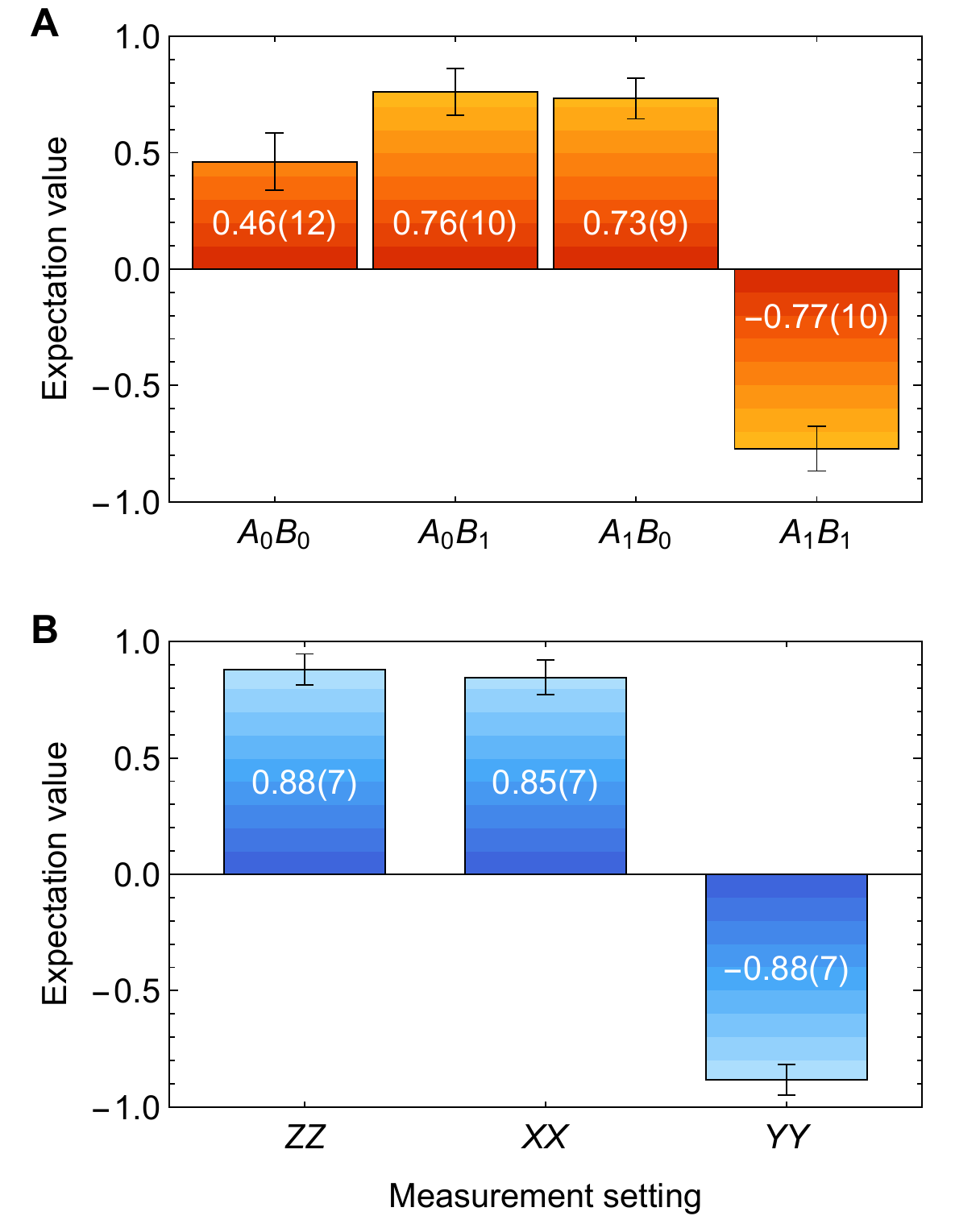}
	 \caption{\textbf{Two-node entanglement.} Results of the Bell test (\textbf{A}) and the correlation measurement (\textbf{B}) with the remote atom-atom entanglement $\ket{\Psi^+_{\text{aa}}}$. Error bars indicate one standard deviation of the photon-counting statistics.}
	\label{fig:entanglement}
\end{figure}

Finally, we run the full entangling scheme and verify the entanglement  when it evolves to a Bell state $\ket{\Psi^{+}_{\mathrm{aa}}}=(\ket{\Downarrow}\ket{U}+\ket{\Uparrow}\ket{D})/\sqrt{2}$. We first measure the $S$ parameter in the Clauser–Horne–Shimony–Holt (CHSH)-type Bell inequality,
\begin{equation}
	S=\left|\braket{A_0\cdot B_0}+\braket{A_0\cdot B_1}+\braket{A_1\cdot B_0}-\braket{A_1\cdot B_1}\right|. \nonumber
\end{equation}
For node A, two observables $A_0$ and $A_1$ are $Z$ and $X$, respectively, and for node B, two observables $B_0$ and $B_1$ are $(-Z+X)/\sqrt{2}$ and $(-Z-X)/\sqrt{2}$, respectively. Results for four settings are shown in Fig.~\ref{fig:entanglement}A. We obtain $S=2.73\pm0.20$, which violates the classical bound of $S\leq 2$ by more than 3 standard deviations. To qualify the entanglement more quantitively, we measure its fidelity with respect to the target state $\ket{\Psi^+_{\mathrm{aa}}}$ as
\begin{equation}
	\mathcal{F}=\text{Tr}(\ket{\Psi_{\mathrm{aa}}^{+}}\bra{\Psi_{\mathrm{aa}}^{+}}\rho)=\frac{1}{4}\left(1+\braket{XX}-\braket{YY}+\braket{ZZ}\right). \nonumber
\end{equation}
Fig.~\ref{fig:entanglement}B summarizes the observed data and we find $\mathcal{F}=0.90\pm0.03$. This fidelity considerably exceeds the threshold of $\mathcal{F}>0.5$ to witness entanglement for a Bell state and step into the regime for practical applications~\cite{wehner2018}.

The overall atom-atom entangling efficiency in each trial is $0.03\%$, including the write-out photon generation efficiency and coupling efficiency, DFG efficiency, fiber transmission efficiency, SFG efficiency, and the EIT mapping-in efficiency. One main limitation lies in the process of atom-photon entanglement generation in node A, since it is intrinsically probabilistic. Exciting the atomic ensemble with write pulses of higher intensity will increase the write-out photon generation efficiency, but simultaneously leads to more contribution from higher-order events and gets entanglement fidelity reduced. One solution is making use of a deterministic scheme of entanglement generation via Rydberg blockade~\cite{sun2021}. Through optical engineering and optimization,  harnessing an atomic ensemble with a very high optical depth for node B, one may push the overall entangling efficiency towards a pure fiber attenuation. Currently, the success of remote entanglement is post-selected via the photon detection at node B, which may give some limitations for potential applications. A better choice will be using a heralded quantum memory~\cite{tanji2009} instead, which will extend the range of applications significantly. The remote entanglement fidelity in our experiment is also limited by the high-order events in node A. Adopting a deterministic scheme with Rydberg atoms will shift this problem to improving the precision of the Rydberg-state manipulations. Our experiment demonstrates the very elementary process of quantum networking at the metropolitan scale, and adoption of similar techniques in a multi-node configuration~\cite{jing2019,pompili2021} will enable functionalities significantly beyond a two-node scenario. 

This work was supported by National Key R\&D Program of China (No.~2017YFA0303902, No.~2020YFA 0309804), Anhui Initiative in Quantum Information Technologies, National Natural Science Foundation of China, and the Chinese Academy of Sciences. We acknowledge QuantumCTek for the allocation of Node A. 

\setcounter{figure}{0}
\setcounter{table}{0}
\setcounter{equation}{0}

\onecolumngrid

\global\long\def\theequation{S\arabic{equation}}
\global\long\def\thefigure{S\arabic{figure}}
\renewcommand{\thetable}{S\arabic{table}}

\newcommand{\msection}[1]{\begin{center} \textbf{#1} \end{center}}

\renewcommand{\thesubsection}{\arabic{subsection}}

\vspace{1.5cm}

\msection{SUPPLEMENTARY MATERIAL}

\section{Details of the experimental setup}
\subsection{General details}
\begin{figure}[!htbp]
    \centering
    \includegraphics[width=0.9\textwidth]{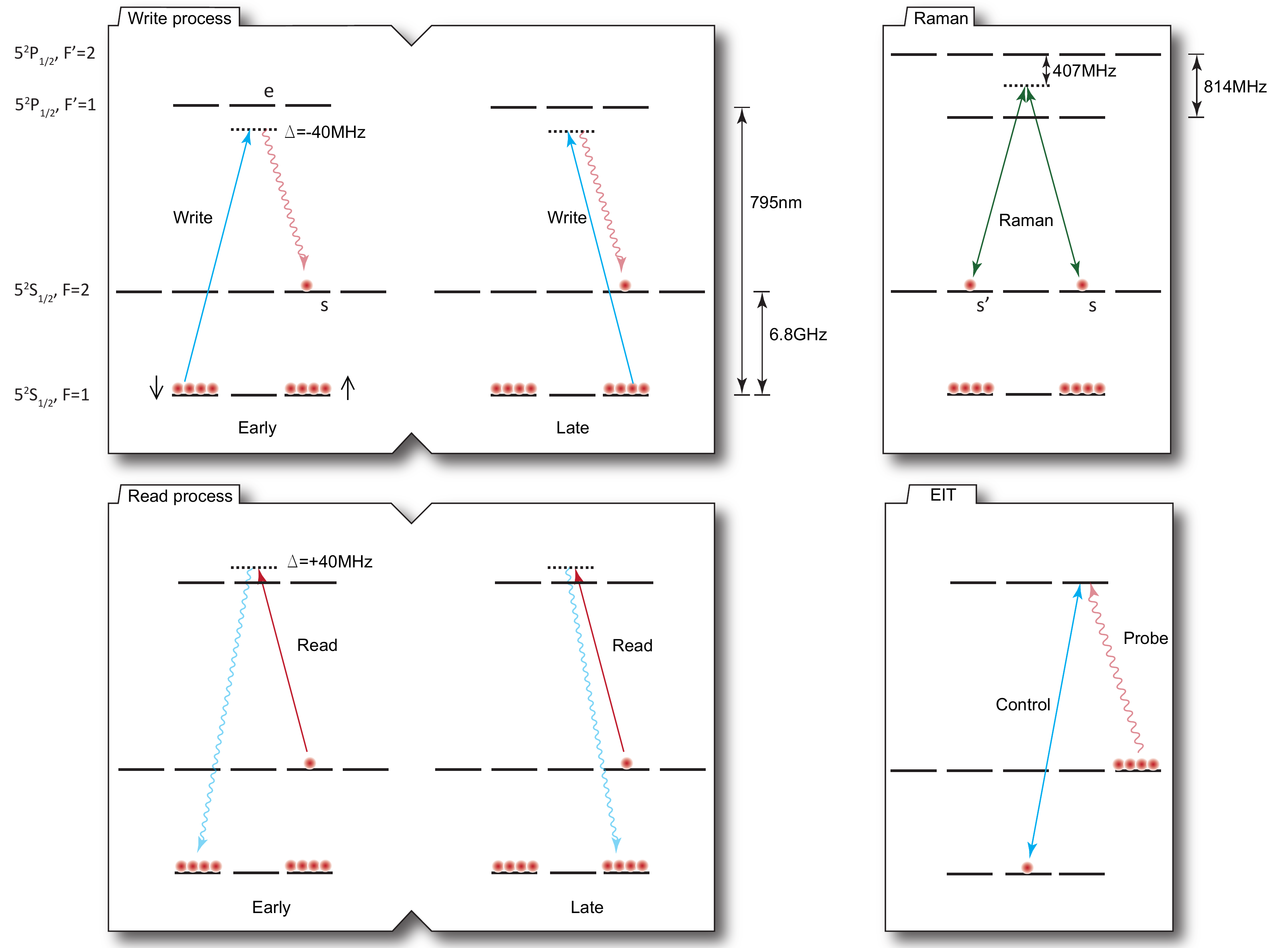}
    \caption{Energy level scheme of important steps in the experiment.}
    \label{fig:energy_scheme}
\end{figure}

Our experiment runs by a repetition rate of $\sim$10~Hz. In each cycle of the experiment, the first 97~ms is for atom preparation, and the rest 3~ms is for running the entanglement distribution scheme. In node A, atom preparation includes 96~ms of atom cooling and trapping by using a standard magneto-optical trap (MOT) and 1~ms of ground state preparation. To prepare atoms equally into $mF=\pm 1$ Zeeman sublevels of the $^{87}$Rb ground states $\ket{5^2S_{1/2},F=1}$, we apply a pumping beam coupling $\ket{5^2S_{1/2},F=1}\leftrightarrow \ket{5^2P_{3/2},F'=0}$ with $\pi$ polarization and a depumping beam coupling $\ket{5^2S_{1/2},F=2}\leftrightarrow \ket{5^2P_{3/2},F'=2}$ to clean atoms staying at $\ket{5^2S_{1/2},F=2}$. In node B, after the first 30~ms of a standard MOT, we switch to the ‘dark SPOT’ configuration to increase the atomic density in the next 66~ms (see Ref.~\cite{yu2021} for details). Fig.~\ref{fig:energy_scheme} gives the energy level scheme of important steps in the experiment, including the write, read process of the DLCZ quantum memory, the Raman operation of the spin-wave freezing, and the storing process of the EIT quantum memory.

\subsection{Control system}
The core of our control system in each node is a field-programmable gate array (FPGA) with several digital inputs and outputs. It runs at a clock frequency of 50~MHz. Experiment sequence control is achieved by switching on/off to the AOMs, which plays the role of fast optical switches. The switching is realized by controlling the radio-frequency sources driving the AOMs, which are homemade direct digital synthesis (DDS) modules in our system. The output of single-photon detectors is sent to the input of the FPGA for data acquisition. Another clock eight times multiplied from the FPGA main clock is used for data acquisition to get a time resolution of 2.5~ns. The raw data in the FPGA buffer will be regularly uploaded to the computer for analysis. 
We set up three classical communication channels with three fibers between two nodes. In channel 1, we use two fiber media converters to run a local area network (LAN). The parameter check and the data sharing are regularly performed in this LAN. In channels 2 and 3, we send the clock frequency and triggers from node A to node B, ensuring an accurate time synchronization between two FPGAs. 

\begin{figure}[htbp]
    \centering
    \includegraphics[width=0.7\textwidth]{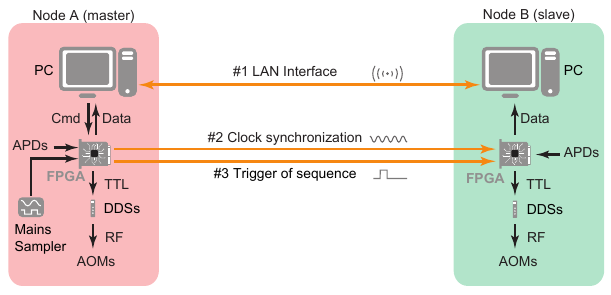}
    \caption{Schematic of the control system}
    \label{fig:remote_scheme}
\end{figure}

\subsection{Laser frequency locking}
Each node has five lasers, two in 780~nm for atom cooling and trapping, two in 795~nm for write, read or control, and one in 1950~nm for frequency conversion. 795~nm and 780~nm lasers are locked to $^{87}$Rb D1 and D2 line, respectively, via Rb vapor cells or high-finesse ultra-stable cavities. One part of the 1950~nm laser is sent to node B for frequency comparison. 

\begin{figure}[htbp]
    \centering
    \includegraphics[width=0.7\textwidth]{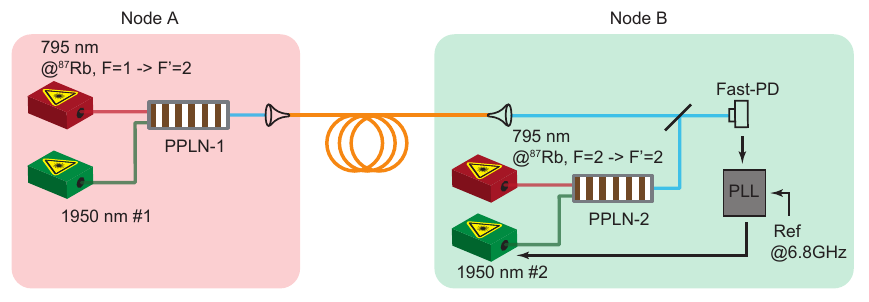}
    \caption{Schematic of the remote frequency locking}
    \label{fig:freq_locking}
\end{figure}

The 1950~nm lasers in two nodes are locked via frequency comparison as the scheme shown in Fig.~\ref{fig:freq_locking}. In each node, we set up an auxiliary QFC module for generating 1342~nm light by combining a 795~nm beam and a 1950~nm beam. The 1342~nm light generated in node A is sent to node B to interfere with the 1342~nm light generated in this node. We send the beat signal to a phase-locking loop (PLL) and feed the feedback signal to the piezo in the pump laser in node B to compensate for the frequency difference. There is a 6.8~GHz frequency difference between the 795~nm lasers in two nodes (laser in node A and B are locked to $F=1 \rightarrow F'=2$ and $F=2 \rightarrow F'=2$ of $^{87}$Rb, respectively). With a 6.8~GHz micro-wave as PLL reference, we lock the pump laser in node B at the same frequency as the pump laser in node A. 

\section{Spin-wave freezing}
In the optimal case, by setting two Raman beams perfectly overlaying the write beam and write-out photon path, one can introduce a wavevector of $\vec{k}_R=-\Delta\vec{k}$ and zero the spin-wave wavevector. In our experiment, the ring cavity at the write-out path makes this configuration difficult to implant. We choose to set one Raman beam that overlays the write beam and the other mirror-symmetric along with the write-out mode. The initial and altered wavevector can be estimated as $|\Delta\vec{k}| \approx 2\pi\cdot \theta /\lambda $ and $|\Delta\vec{k}'| \approx |\Delta\vec{k}|\cdot \theta$, where $\lambda=795$~nm is the wavelength and $\theta=3.5^{\circ}$ is the angle between write and write-out. With the knowledge of the atomic temperature $T\approx 35~\upmu$K, we can estimate the expected lifetime in the spin-wave freezing case $\tau = 586~\upmu$s. 

\section{Cancellation of the phase noise}
The magnetic fields for controlling atoms in the experiment are generated by Helmholtz coils driven by DC sources. Magnetic noise introduced in this process will lead to qubit dephasing. Specifically, the noise is introduced in two ways. First, the transformers or AC/DC converters inside the DC sources will generate magnetic fields and be felt by atoms. Second, The residual AC noise in the DC sources output will generate AC magnetic field via Helmholtz coils.

\begin{figure}
	\begin{minipage}[b]{.45\textwidth}
	  \centering
	  \includegraphics[width=\textwidth]{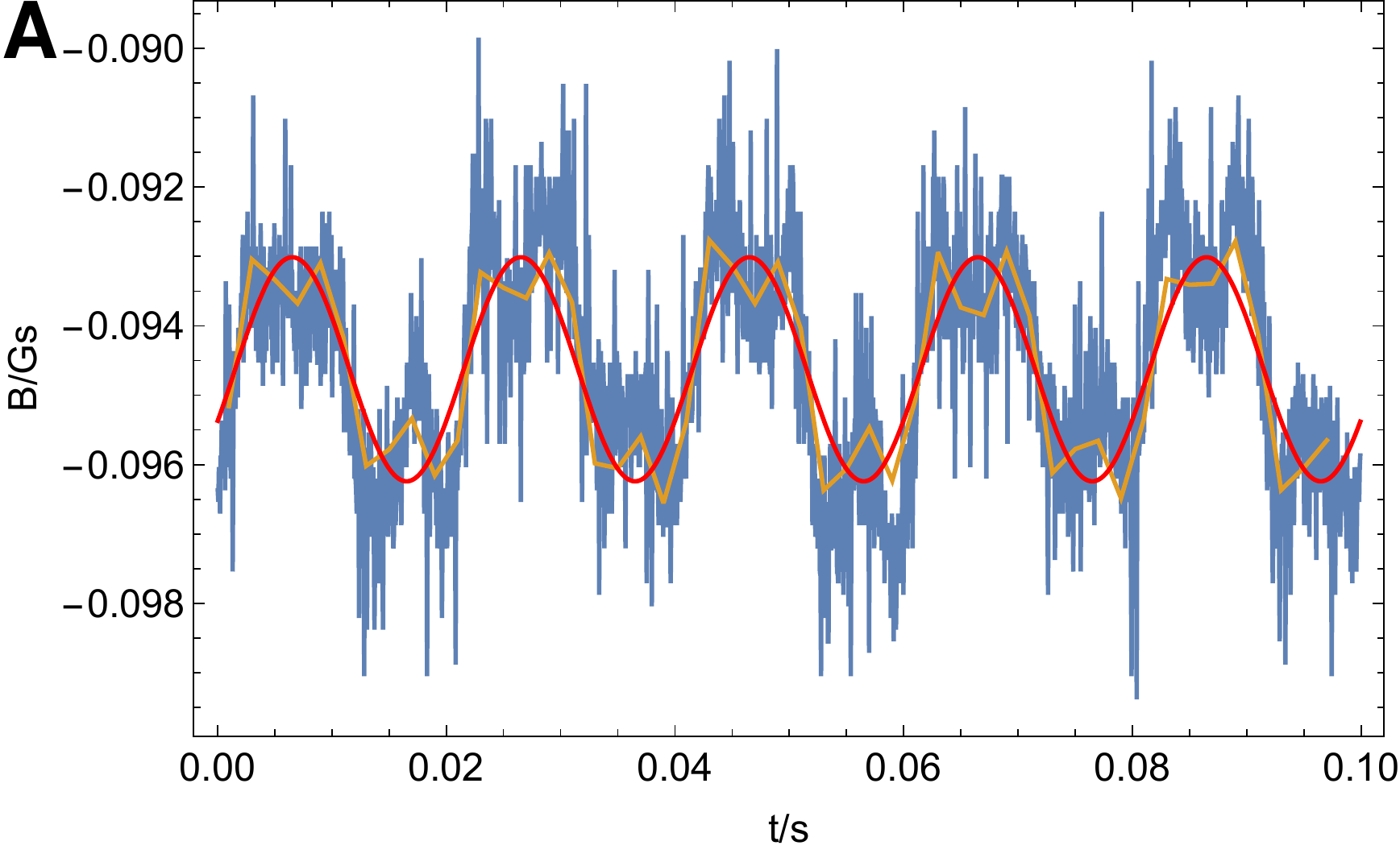}
	\end{minipage}
	\begin{minipage}[b]{.45\textwidth}
	  \centering
	  \includegraphics[width=\textwidth]{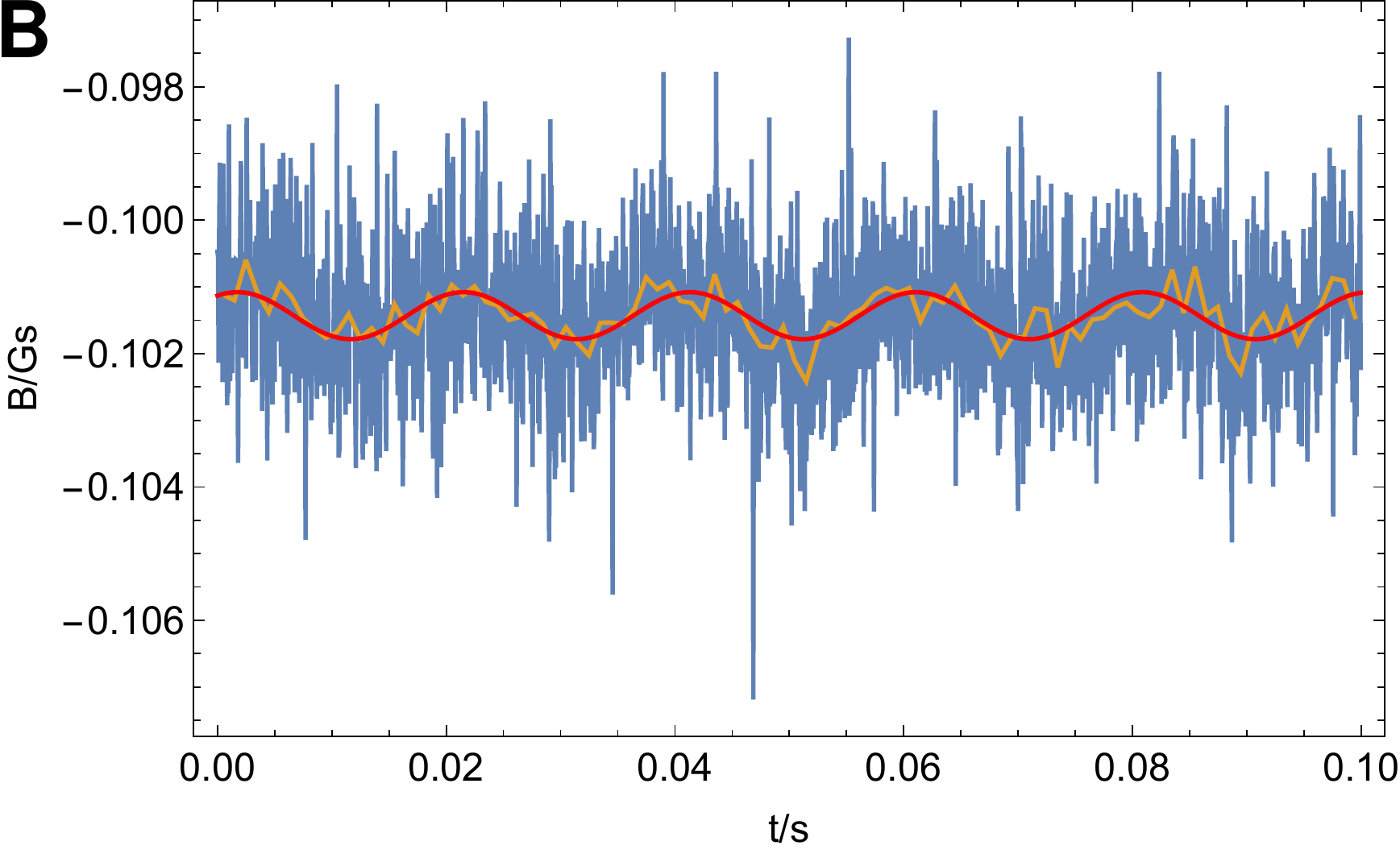}
	\end{minipage}
	\caption{Magnetic field at 0.5~m (A) and 3~m (B) away from the DC sources as a function of time. Blue points are the raw data, and the orange line shows an average of the data with a time-bin of 2~ms. The red line is a sinusoidal fit of the orange data.}
	\label{fig:magnetic_noise}
\end{figure}

\begin{figure}
	\centering
	\includegraphics[width=.5\textwidth]{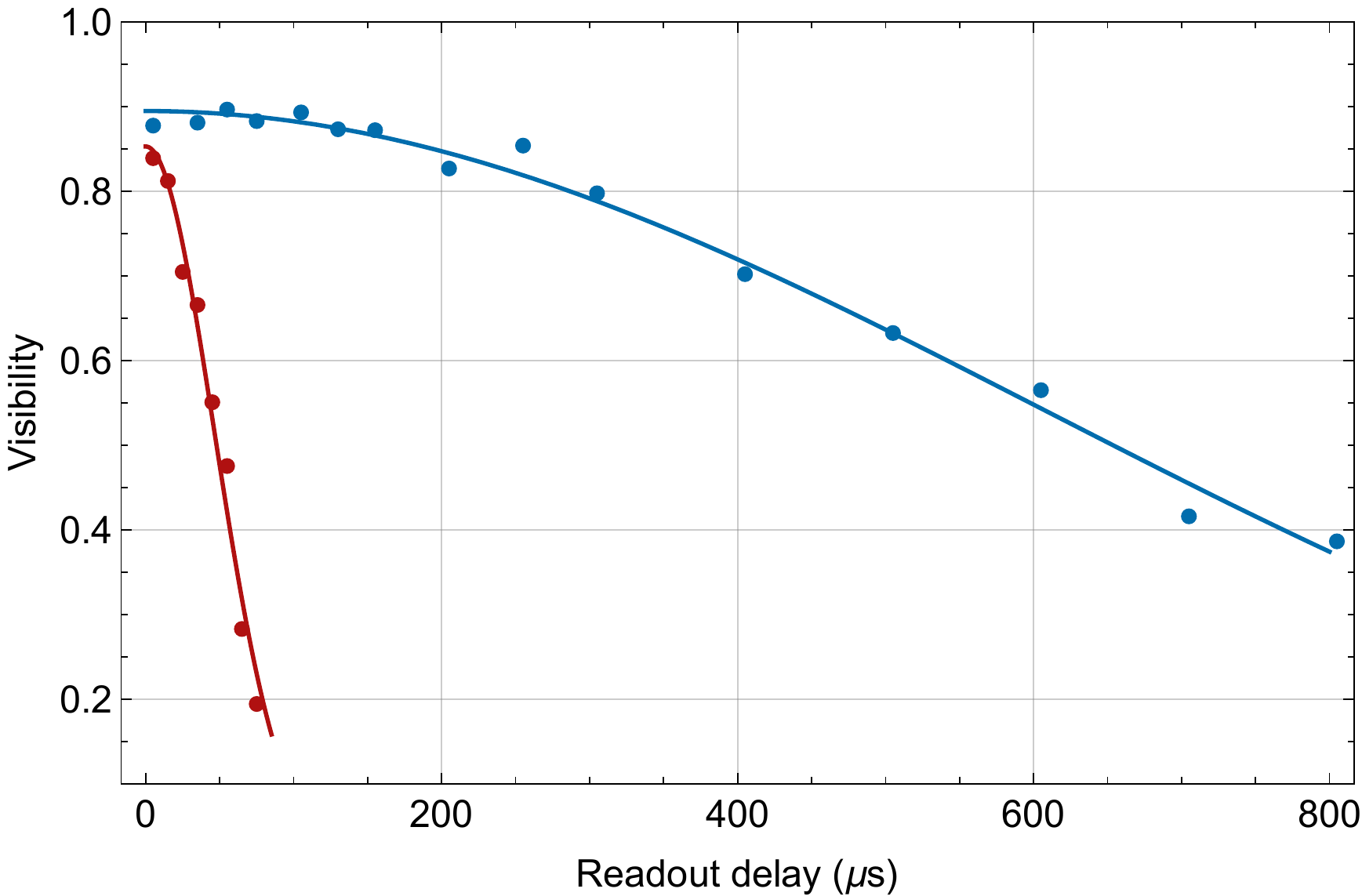}
	\caption{Atom-photon correlation as a function of the readout delay. Red (blue) points show the visibility of the correlation in $XX$ basis before (after) the phase noise cancellation.}
	\label{fig:visibility_compare}
\end{figure}

Fig.~\ref{fig:magnetic_noise} shows the magnetic field measured 0.5~m and 3~m away from the DC sources. We observed a 50~Hz oscillation in both cases, but the oscillation amplitude decreased from 1.61~mG to 0.35~mG. To isolate this noise, we keep a distance of $>3$~m between the DC sources and the atomic cell. Given that the noise is periodical and its 20~ms period is much larger than 3~ms of the entanglement scheme running time, we synchronize the experiment sequence with the Mains electricity frequency to further eliminate its influence. We use a Mains frequency sampler to sample the 220~V sine wave Mains electricity to a 3~V square wave reference signal and synchronize the experiment sequence with the reference signal. Fig.~\ref{fig:visibility_compare} shows the correlation of the write-out photon and the spin-wave on the $XX$ basis. The fitting shows that $T_2^*$ is increased from $65.5~\mu$s to $856.7~\mu$s after the noise cancellation measures are applied.

\section{Atom-atom entangling efficiency}
The final atom-atom entanglement is measured via converting to a pair of photon-photon entanglement. The overall photon-photon coincidence probability measured is $6.1\times10^{-6}$. The readout efficiency is 0.15 for node A, and 0.13 for node B, including the atom-to-photon mapping efficiency, coupling efficiency, transmission efficiency and detector efficiency. By correcting the two readout efficiencies, we get an atom-atom entangling efficiency of 0.03\%. 

\bibliography{myref}

\end{document}